\documentclass[prc,twocolumn,showpacs,amsmath,amssymb]{revtex4}
\usepackage{graphicx}
\usepackage{dcolumn}
\usepackage{bm}
\begin{document}

\title{Coupled-channels density-matrix approach to low-energy nuclear reaction dynamics}

\author{Alexis Diaz-Torres}

\affiliation{Department of Physics, University of Surrey, Guildford, GU2 7XH, 
United Kingdom}

\date{\today}

\begin{abstract}
Atomic nuclei are complex, quantum many-body systems whose structure manifests 
itself through intrinsic quantum states associated with different excitation modes or
degrees of freedom. Collective modes (vibration and/or rotation) dominate at low
energy (near the ground-state). The associated states are usually employed, within a truncated model space, as a basis in (coherent) coupled channels approaches to low-energy reaction dynamics. However, excluded states can be essential, and 
their effects on the open (nuclear) system dynamics are usually treated through
complex potentials. Is this a complete description of open system dynamics? Does it include effects of quantum decoherence? Can decoherence be manifested in reaction
observables? In this contribution, I discuss these issues and the main ideas of a 
coupled-channels density-matrix approach that makes it possible to quantify the role 
and importance of quantum decoherence in low-energy nuclear reaction dynamics. Topical applications, which refer to understanding the astrophysically important collision 
$^{12}$C + $^{12}$C and achieving a unified quantum dynamical description of relevant reaction processes of weakly-bound nuclei, are highlighted.  
\end{abstract}


\maketitle

\section{Introduction} 

Nuclear reaction research has entered a new era with developments of radioactive ion beam facilities, at which nuclear reactions are the primary probe of the new physics, such as novel structural changes, through dynamical excitations of nucleonic, collective and cluster degrees of freedom. Innovative detection systems are allowing measurements of unprecedented exclusivity and precision, including those using intense stable beams. These and the increased intensity rare-isotope beam capabilities require investigations of the role of hitherto innaccessible degrees of freedom and new considerations in quantum nuclear dynamics. Properly combining reaction dynamics and many-body structure information is a frontier research area across different fields of physics and chemistry. 

\begin{figure}
  \includegraphics[width=0.45\textwidth,angle=0]{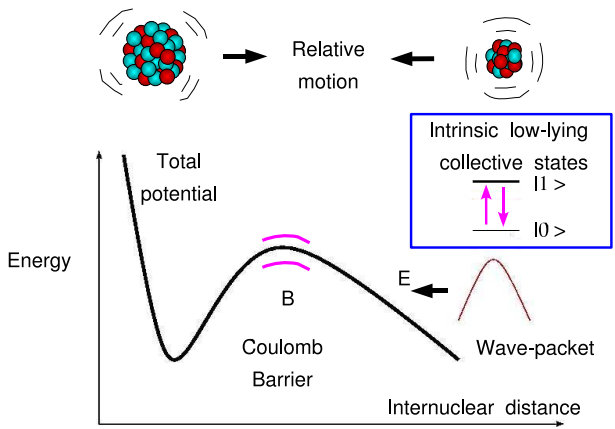}
  \newline
  \caption{Schematic of the coherent coupled-channels description of 
a low-energy collision.}
\label{Figure1}
\end{figure}

The coherent quantum dynamics of a low-energy nuclear collision is illustrated in 
Fig.\ \ref{Figure1}. The two interacting nuclei are initially at the ground states, but get intrinsically excited -as they approach- due to the mutual Coulomb and nuclear interactions. These two opposite interactions result in a bare total potential that has a Coulomb barrier. A wave-packet with certain average incident energy describes the relative motion. It is usually considered to be coupled to a few intrinsic, low-lying collective states that keep their quantum phase relationship, the dynamics being dictated by the Schr\"odinger equation. This leads to a coherent quantum superposition of intrinsic states that has a major consequence: the bare Coulomb barrier splits into individual barriers associated with the specific, intrinsic quantum states (fusion barriers distribution \cite{Nanda1}). These determine different fusion pathways that interfere with each other. Fusion happens when these barriers are overcome, and the nuclei are irreversibly trapped in the potential pocket inside the barriers. In general, the coherent quantum superposition of intrinsic states enhances the total fusion probability \cite{Dasso}, compared to the probability for the (single) bare potential barrier. Crucially, this can be tested against high-precision fusion measurements.   

New, precise fusion measurements \cite{Nanda2,David1} over the last few years have systematically shown disagreement with predictions of the coherent coupled-channels picture. It has also failed in describing the elastic and quasi-elastic scattering and fusion processes simultaneously \cite{Newton1}. This has inevitably led to phenomenological (sometimes contradictory) adjustments \cite{Esbensen1,Hagino1} to stationary-state coupled channels models to fit the experimental data, but without a physically consistent foundation. 

With collaborators, I have suggested \cite{Alexis1} that quantum decoherence and energy dissipation should be simultaneously included in a consistent description of low-energy reaction dynamics. A possible description is the coupled-channels density-matrix ({\sc ccdm}) approach \cite{Alexis1,Alexis2}. A survey of theoretical approaches to dissipative dynamics of low-energy nuclear collisions is provided in Ref. \cite{Alexis1}. Most of these developments, in contradistinction to the {\sc ccdm} approach, do \emph{not} treat the relative motion of the nuclei quantum-mechanically and/or use \emph{incoherent} (statistically averaged) rather than \emph{decoherent} (partially coherent) reaction channels. This paper discusses the main ideas of the {\sc ccdm} approach and highlights two topical applications.

\section{COUPLED CHANNELS DENSITY MATRIX APPROACH}

\begin{figure}
%\begin{center}
%\begin{tabular}{cc}
\includegraphics[width=0.50\textwidth,angle=0]{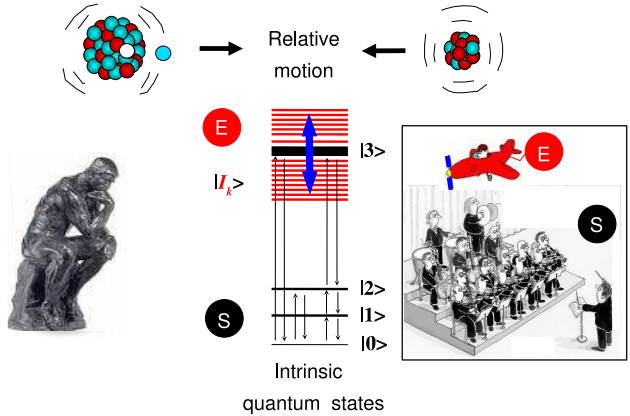}
%\end{tabular}
\caption{A low-energy collision represented by an open quantum system (relative motion + a few intrinsic, low-lying collective states). The high density of single-particle states surrounding a giant resonance state represents the environment. It gradually destroys the coherent quantum superposition of the reduced-system collective states, as the nuclei approach.}
\label{Figure2}
%\end{center}
\end{figure}

Figure\ \ref{Figure2} shows the key ideas of the {\sc ccdm} approach. The reduced 
quantum system comprises the relative motion of the nuclei and a few intrinsic, low-lying collective states, whilst the bath of nucleonic excitations represents the environment. It significantly affects the dynamics of the reduced quantum system. What are these effects?

To answer this question, I will use the cartoons in Fig.\ \ref{Figure2}. The reduced system is represented by the orchestra, where the director plays the role of the relative motion and the musicians correspond to the selected, collective states. The airplane represents the environment. When the airplane is not present, the orchestra plays a wonderful music, all musicians are in sync, like in a coherent quantum superposition. But when the airplane appears, the listener percieves two effects due to interference: (i) the music gets attenuated (dissipation/absorption) and, most importantly, (ii) the quality of the music changes, as the musicians play out of sync (decoherence). 

\begin{figure}[h]
%\begin{center}
\includegraphics[width=0.45\textwidth,angle=0]{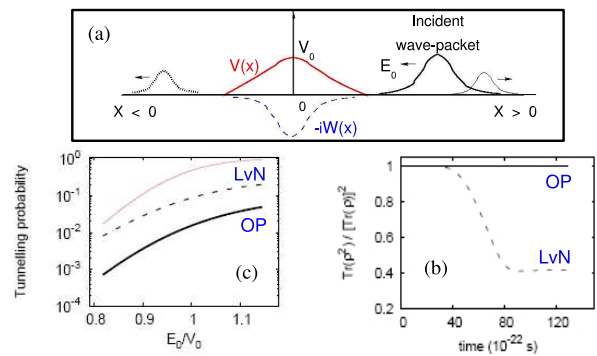}
\caption{(a) One-dimensional model of a wave-packet scattering off a potential barrier \cite{Alexis3}. Dynamical calculations, using the optical potential model (OP) and the Lindblad dynamics (LvN), are carried out for (b) a measure of coherence and (c) the quantum tunnelling probability. Clearly, the two descriptions are \emph{not} equivalent.} 
\label{Figure3}
%\end{center}
\end{figure}

The latter really makes these ideas, applied to low-energy nuclear collisions, innovative. This is because in the widely used optical model for nuclear reactions, where a complex potential describes the effects of excluded degrees of freedom, \emph{only} absorption is described. The absence of quantum decoherence in the complex-potential approach to nuclear scattering has recently been demonstrated in Ref. \cite{Alexis3}, within a simple model illustrated in Fig.\ \ref{Figure3}(a). Here, a wave-packet scatters off a potential barrier, and dynamical calculations are carried out for a measure of the spatial coherence [Fig.\ \ref{Figure3}(b)] and the quantum tunnelling probability [Fig.\ \ref{Figure3}(c)]. Whilst the optical model preserves coherence [thick solid line in Fig.\ \ref{Figure3}(b)], the Lindblad dynamics results in loss of coherence (dashed line). Clearly, the two descriptions are \emph{not} equivalent, and the impact of decoherence on the tunnelling probability is quite substantial [comparing the thick solid to the dashed line in Fig.\ \ref{Figure3}(c)]. It is also observed, comparing these two lines to the thin solid line representing the tunnelling probability without environmental effects, that decoherence changes the energy dependence of the tunnelling probability significantly. I conclude that a deterministically evolving wave-function (pure state) cannot describe quantum decoherence which is a dynamical process where a pure state becomes a mixed state. A description based on either a time-dependent density matrix or an ensemble of stochastically evolving wave functions (Monte Carlo wave-function method \cite{Molmer}) is essential for quantifying quantum decoherence effects on reaction observables. 

The {\sc ccdm} approach is based on the Liouville-von Neumann master equation with Lindblad dissipative terms \cite{Alexis1,Alexis2}. This technique was first introduced in studies of quantum molecular dynamics \cite{Saalfrank}. The Lindblad terms consistently account for dissipation and decoherence. The crucial idea is to project this master equation onto a product-state basis, where a part of the basis describes the internuclear separations (grid-basis) and another part describes the selected, intrinsic (collective) states of the interacting nuclei. This yields a finite set of coupled equations for the time-dependent density-matrix elements. The initial density matrix is clearly determined when the two nuclei are well-separated at the ground states and a Gaussian wave-packet describes the radial motion. I have developed the formalism presented in Refs. \cite{Alexis1,Alexis2} further, using a coupled-angular-momentum state basis. This is very useful for investigating decoherence effects on asymptotic observables, such as the angular distribution of inelastic excitations. Results will be reported elsewhere.

For the sake of simplicity, I referred to only one specific environment in Fig.\ \ref{Figure2}, i.e., the sea of nucleonic excitations surrounding a giant resonance state of a colliding nucleus. However, various types of environments can coexist in a nuclear collision, which may be specific to particular degrees of freedom, such as weak binding or isospin asymmetry. Among these environments, which can be coupled to specific states or to all states of the reduced system, are (i) the multitude of one- and multi-nucleonic excitations in mass/charge partitions other than the entrance one (transfer), (ii) the continuum of non-resonant decay states of weakly-bound nuclei (breakup), and (iii) the innumerable nuclear molecular (compound nucleus) states (fusion). These can be treated separately, and their effects can be distinguished within the {\sc ccdm} approach.

\begin{figure}[h]
\includegraphics[width=0.45\textwidth,angle=0]{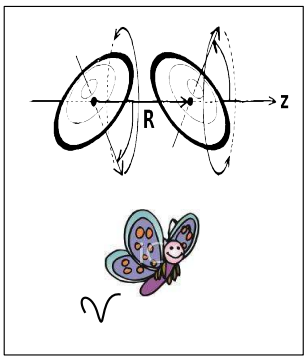}
\caption{A nuclear molecule of two oblately deformed nuclei, exhibiting complex 
excitation modes, like a butterfly flapping its wings.} 
\label{Figure4}
\end{figure}

\section{APPLICATIONS}

The {\sc ccdm} approach finds a wide range of applications in areas of low-energy nuclear reaction physics. For instance, an excellent unresolved problem is understanding fusion of astrophysically important collisions like $^{12}$C + $^{12}$C \cite{Aguilera,Spillane}. Of relevance here is to know the fusion probability at energies near the Gamow peak ($\sim$ 1.5 MeV). It is usually obtained from extrapolations of high-energy fusion measurements, as direct experiments are extremely difficult to carry out at very low incident energies (< 3 MeV). The presence of pronounced resonance structures in the fusion excitation function makes it quite uncertain. Understanding the origin of these resonances and their impact on the reaction rates is a long-standing problem in heavy-ion physics. 

\begin{figure}[h]
\begin{tabular}{cc}
\includegraphics[width=0.25\textwidth,angle=0]{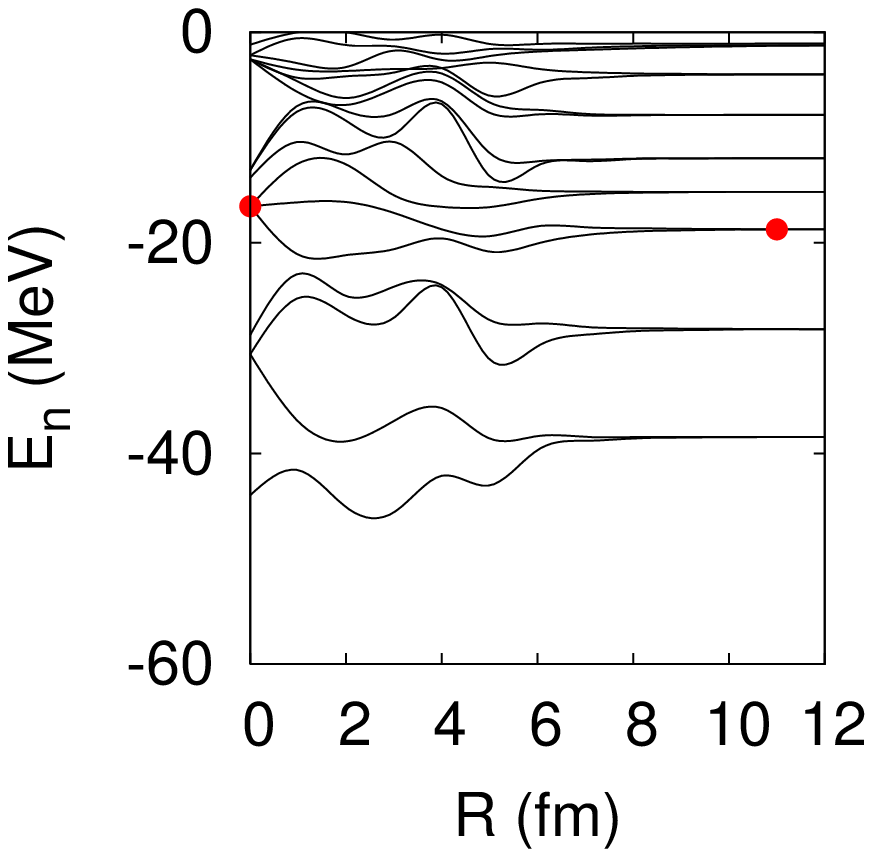} &
\includegraphics[width=0.25\textwidth,angle=0]{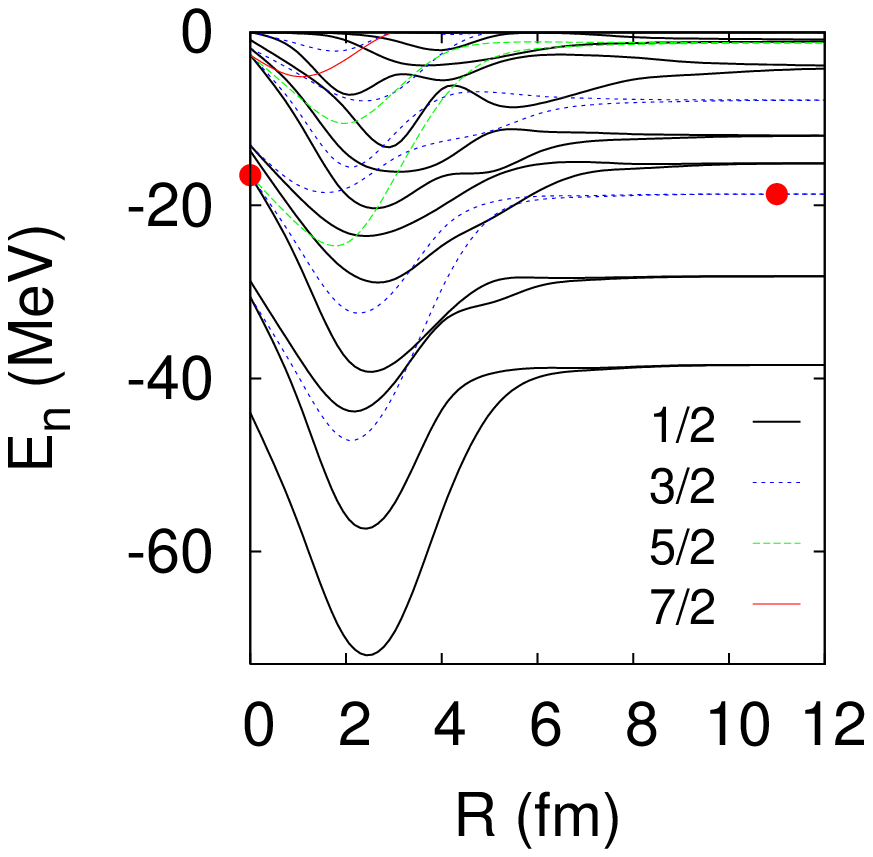} \\
\includegraphics[width=0.15\textwidth,angle=0]{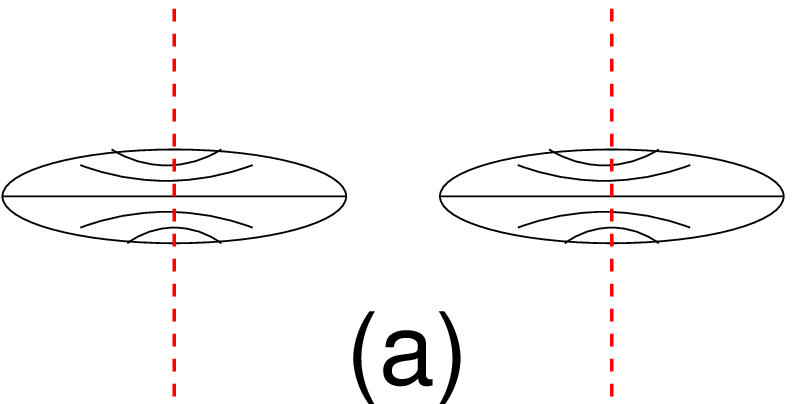} &
\includegraphics[width=0.15\textwidth,angle=0]{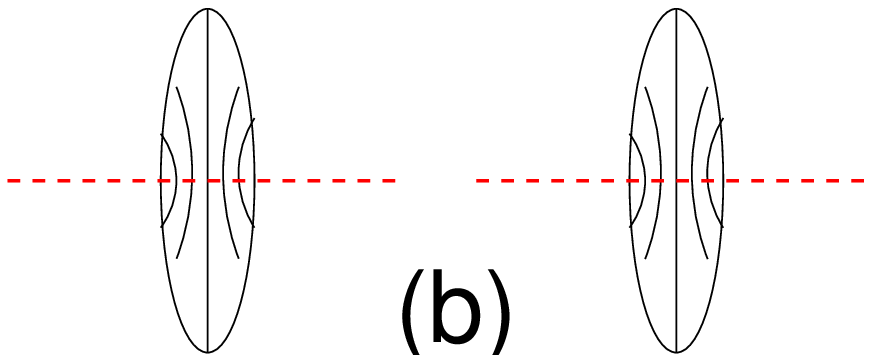}
\end{tabular}
\caption{Neutron molecular shell structure of two interacting $^{12}$C nuclei as a function of the internuclear distance \cite{Alexis6} for configurations: (a) non-axial symmetric and (b) axial symmetric, where the different lines denote magnetic sub-states. The spectrum at small distances is very sensitive to the nuclei alignment.} 
\label{Figure5}
\end{figure}

The resonances may be mainly related to complex excitation modes in the dinuclear system, when the two $^{12}$C nuclei come into contact, as illustrated in Fig.\ \ref{Figure4}. The $^{12}$C intrinsic symmetry axis vibrates and/or rotates with respect to the internuclear axis. These molecular dynamical modes are opened up when the nuclei overlap, and supply a complex environment. It can decohere the rotational states of the separated, individual 
$^{12}$C nuclei, which are excited by the long-range Coulomb mechanism. Using a realistic two-center shell model \cite{Alexis5,Alexis6}, I have demonstrated \cite{Alexis6} that the single-particle molecular shell structure at small internuclear distances is very sensitive to the alignment of the $^{12}$C nuclei (see Fig.\ \ref{Figure5}). Non-axial symmetric configurations preserve the individuality of the overlapping nuclei (the asymptotic shell structure is largely maintained) [Fig.\ \ref{Figure5}(a)], whilst this is not the case for the axial symmetric configuration [Fig.\ \ref{Figure5}(b)]. The former favors re-separation, and the latter fusion. The competition among these configurations, as a function of the incident energy and orbital angular momentum, should result in molecular resonance structures in the fusion excitation function.

\begin{figure}[h]
\includegraphics[width=0.50\textwidth,angle=0]{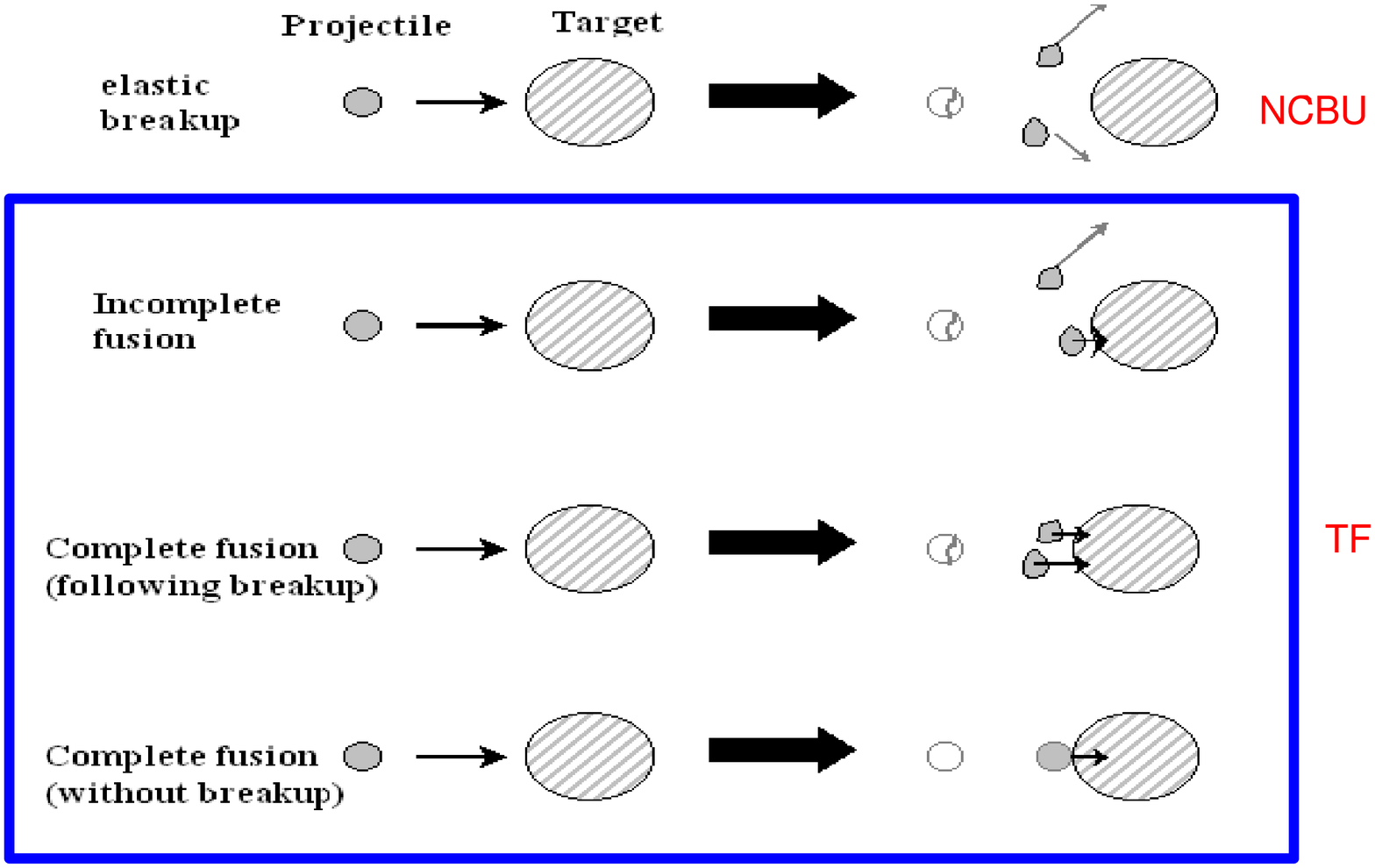}
\caption{Schematic of some relevant reaction processes of a two-body, weakly-bound projectile colliding with a stable target. The no-capture breakup ({\sc ncbu}) process and different components of the total fusion ({\sc tf}) process are highlighted. 
} 
\label{Figure6}
\end{figure} 

Another great theoretical challenge is to achieve a unified quantum dynamical description of relevant reaction processes of weakly-bound nuclei. Some of these are presented in Fig.\ \ref{Figure6}. Although it is not illustrated there, the transfer process is also very important \cite{Navin,Shrivastava,Chatterjee,Ramin}. Existing quantum models have limitations \cite{Thompson}, as they cannot calculate integrated incomplete and complete fusion cross sections unambiguously \cite{Alexis9,Alexis10}. Neither, after the formation of incomplete fusion products, can they follow the evolution of the surviving breakup fragment(s) since incomplete fusion results in depletion of the total few-body wave-function. Some difficulties are overcome by the classical dynamical model suggested in Refs. \cite{Alexis7,Alexis8}. However, a quantum model is very desirable, as it can deal with quantum tunnelling that is essential for understanding astrophysical reaction rates involving exotic nuclei. One possibility of tackling this issue is through the {\sc ccdm} approach.     

\section{SUMMARY}

The main ideas of an innovative, coupled-channels density-matrix approach to low-energy nuclear reaction dynamics have been presented. It will quantify the role and importance of quantum decoherence in various areas of nuclear reaction theory, interlacing nuclear structure information and reaction physics. Decoherence should always be explicitly included when modelling low-energy reaction dynamics with a limited set of (relevant) degrees of freedom. 

\begin{acknowledgments}
Useful discussions with participants in the ECT$^*$ Workshop on Decoherence in Quantum Dynamical Systems, Trento, April 26--30, 2010, are acknowledged. The work was supported 
by the UK Science and Technology Facilities Council (STFC) Grant No. ST/F012012/1. 
\end{acknowledgments}


\begin{thebibliography}{99}
\bibitem{Nanda1} M.~Dasgupta et al., Annu. Rev. Nucl. Part. Sci. \textbf{48}, 401 (1998).
\bibitem{Dasso} C.~H.~Dasso, S.~Landowne and A.~Winther, Nucl. Phys. A \textbf{432}, 495 (1985).
\bibitem{Nanda2} M.~Dasgupta et al., Phys. Rev. Lett. \textbf{99}, 192701 (2007), and references therein.
\bibitem{David1} D.~Hinde et al., Nucl. Phys. A \textbf{834}, 117c (2010).
\bibitem{Newton1} J.~O.~Newton et al., Phys. Rev. C \textbf{70}, 024605 (2004).
\bibitem{Esbensen1} S.~Misicu and H.~Esbensen, Phys. Rev. Lett. \textbf{96}, 112701 (2006).  
\bibitem{Hagino1} T.~Ichikawa, K.~Hagino and A.~Iwamoto, Phys. Rev. Lett. \textbf{103}, 202701 (2009).
\bibitem{Alexis1} A.~Diaz-Torres et al., Phys. Rev. C \textbf{78}, 064604 (2008).
\bibitem{Alexis2} A.~Diaz-Torres et al., AIP Conf. Proc. \textbf{1098}, 44 (2009).
\bibitem{Alexis3} A.~Diaz-Torres, Phys. Rev. C \textbf{81}, 041603(R) (2010).
\bibitem{Molmer} K.~M{\o}lmer, Y.~Castin and J.~Dalibard, J. Opt. Soc. Am. B \textbf{10}, 
524 (1993).
\bibitem{Saalfrank} L.~Pesce and P.~Saalfrank, Chem. Phys. \textbf{219}, 43 (1997); 
J. Chem. Phys. \textbf{108}, 3045 (1998).
\bibitem{Aguilera} E.~F.~Aguilera et al., Phys. Rev. C \textbf{73}, 064601 (2006).
\bibitem{Spillane} T.~Spillane et al., Phys. Rev. Lett. \textbf{98}, 122501 (2007).
\bibitem{Alexis5} A.~Diaz-Torres and W.~Scheid, Nucl. Phys. A \textbf{757}, 373 (2005).
\bibitem{Alexis6} A.~Diaz-Torres, Phys. Rev. Lett. \textbf{101}, 122501 (2008).
\bibitem{Navin} A.~Navin et al., Phys. Rev. C \textbf{70}, 044601 (2004).
\bibitem{Shrivastava} A.~Shrivastava et al., Phys. Lett. B \textbf{633}, 463 (2006).
\bibitem{Chatterjee} A.~Chatterjee et al., Phys. Rev. Lett. \textbf{101}, 032701 (2008).
\bibitem{Ramin} R.~Rafiei et al., Phys. Rev. C \textbf{81}, 024601 (2010). 
\bibitem{Thompson} I.~J.~Thompson and A.~Diaz-Torres, Prog. Theor. Phys. Suppl. \textbf{154}, 69 (2004).
\bibitem{Alexis9} A.~Diaz-Torres and I.~J.~Thompson, Phys. Rev. C \textbf{65}, 024606 (2002). 
\bibitem{Alexis10} A.~Diaz-Torres, I.~J.~Thompson and C.~Beck, Phys. Rev. C \textbf{68}, 044607 (2003). 
\bibitem{Alexis7} A.~Diaz-Torres et al., Phys. Rev. Lett. \textbf{98}, 152701 (2007).
\bibitem{Alexis8} A.~Diaz-Torres, J. Phys. G: Nucl. Part. Phys. \textbf{37}, 075109 (2010).

\end{thebibliography}
\end{document}